\documentclass[twocolumn,superscriptaddress,prl]{revtex4-1}
\usepackage{amssymb}
\usepackage{amsmath}
\usepackage{graphicx,float,calc}
\usepackage{color,bm}
\usepackage[colorlinks,urlcolor=blue,citecolor=blue,linkcolor=blue]{hyperref}
\usepackage{epstopdf}
\usepackage{amsfonts}
\begin{document}

\title{Anomalous symmetry protected blockade of skin effect in one-dimensional non-Hermitian lattice systems}

\author{Shuai Li}
\thanks{These authors contribute equally to this work.}
\affiliation{Ministry of Education Key Laboratory for Nonequilibrium Synthesis and Modulation of Condensed Matter,Shaanxi Province Key Laboratory of Quantum Information and Quantum Optoelectronic Devices, School of Physics, Xi'an Jiaotong University, Xi'an 710049, China}

\author{Min Liu}
\thanks{These authors contribute equally to this work.}
\affiliation{Ministry of Education Key Laboratory for Nonequilibrium Synthesis and Modulation of Condensed Matter,Shaanxi Province Key Laboratory of Quantum Information and Quantum Optoelectronic Devices, School of Physics, Xi'an Jiaotong University, Xi'an 710049, China}

\author{Yue Zhang}
\affiliation{Ministry of Education Key Laboratory for Nonequilibrium Synthesis and Modulation of Condensed Matter,Shaanxi Province Key Laboratory of Quantum Information and Quantum Optoelectronic Devices, School of Physics, Xi'an Jiaotong University, Xi'an 710049, China}

\author{Rui Tian}
\affiliation{Ministry of Education Key Laboratory for Nonequilibrium Synthesis and Modulation of Condensed Matter,Shaanxi Province Key Laboratory of Quantum Information and Quantum Optoelectronic Devices, School of Physics, Xi'an Jiaotong University, Xi'an 710049, China}

\author{Maksims Arzamasovs}
\affiliation{Ministry of Education Key Laboratory for Nonequilibrium Synthesis and Modulation of Condensed Matter,Shaanxi Province Key Laboratory of Quantum Information and Quantum Optoelectronic Devices, School of Physics, Xi'an Jiaotong University, Xi'an 710049, China}

\author{Bo Liu}
\email{liubophy@gmail.com}
\affiliation{Ministry of Education Key Laboratory for Nonequilibrium Synthesis and Modulation of Condensed Matter,Shaanxi Province Key Laboratory of Quantum Information and Quantum Optoelectronic Devices, School of Physics, Xi'an Jiaotong University, Xi'an 710049, China}

\begin{abstract}
The non-Hermitian skin effect (NHSE), an anomalous localization behavior of the bulk states, is an inherently non-Hermitian phenomenon, which can not find a counterpart in Hermitian systems. However, the fragility of NHSE has been revealed recently, such as the boundary sensitivity, and it stimulates a lot of studies on discussing the fate of that. Here we present a theorem which shows that the combined spatial reflection symmetry can be considered as a criterion in one-dimensional non-Hermitian systems to determine whether the NHSE can exist or not. Distinct from previous studies, our proposed criterion only relies on analyzing the symmetry of the system, freeing out other requirements, such as the information of the energy spectrum. Furthermore, by taking the non-Hermitian Kitaev chain as an example, we verify our theorem through both a mathematical proof via the non-Bloch band theory and the exact diagonalization numerical studies. Our results reveal a profound connection between the symmetry and the fate of NHSE.
\end{abstract}

\maketitle
In recent years, non-Hermitian physics~\cite{Hatano1996,Bender1998,Bender2002,Bender2007,Moiseyev2011,ElGanainy2018} has attracted tremendous interests in both theoretical and experimental studies in various systems, such as in non-equilibrium open systems~\cite{Zhen2015,Wu2019,Franca2022,Li2023}, and systems with finite-lifetime quasiparticles~\cite{Papaj2019,Shen2018,Shen2023,Padhan2024}. Various interesting phenomena, including non-Hermitian skin effect (NHSE)~\cite{Yao2018,Zhang2021}, exceptional points (EPs)~\cite{Choi2010,Wang2024}, and non-Hermitian spintronics have been unveiled~\cite{Okuma2019,Rivero2023,Wetter2023}. In particular, NHSE is one of the unexpected effect in non-Hermitian systems, which is mainly manifested by the appearance of the abnormal localization of multitudinous bulk modes when considering under the open boundary condition (OBC)~\cite{Lee2016,Yao2018,Yao2018a,Song2019}. It has been observed in a variety of experimental platforms, including mechanical metamaterials~\cite{Brandenbourger2019}, topolectric circuit~\cite{Helbig2020}, photonic lattice~\cite{Weidemann2020,Zhou2023} and quantum gases~\cite{Liang2022}. To comprehensively understand the associated breakdown of conventional bulk-boundary correspondence, the non-Bloch band theory has been established~\cite{Yao2018,Yao2018a,Yokomizo2019,Yang2020,Zirnstein2021,Zhang2023}. Lots of interesting effects can arise from NHSE, such as unconventional reflection~\cite{Li2022} and entanglement suppression~\cite{Lu2021,Kawabata2023,Gliozzi2024}. Ultra-sensitivity quantum sensors~\cite{Budich2020,Hokmabadi2019,DeCarlo2022}, exponential signal enhancement~\cite{Koch2022,McDonald2020} and unidirectional transport~\cite{Longhi2015a,Du2020} have been predicted.
Recently, there are a lot of studies on discussing the fate of NHSE~\cite{Zhao2021,Yokomizo2021}, including the sensitivity of bulk modes~\cite{Kunst2018,Yi2020,Zhang2022}, influence of lattice size and boundary perturbations~\cite{Guo2021} and topological effect of the point-gap~\cite{Okuma2020,Zhang2020}.

In this work, we develop a symmetry-based  method to examine the origin/reason for the occurrence of NHSE. Distinct from previous studies~\cite{Kunst2018,Yi2020,Zhang2022,Guo2021,Okuma2020,Zhang2020}, our proposed criterion only relies on analyzing the symmetry of the system, freeing out other requirements, such as the information of the energy spectrum~\cite{Okuma2020,Zhang2020}. As we know, symmetry plays a pivotal role in non-Hermitian systems \cite{Bender1998,Mostafazadeh2002,Budich2019,Delplace2021,Mandal2021,Gardas2016}. To build the connection between symmetry and the occurrence of NHSE, here we pay special attention to the spatial reflection ($\mathcal{P}$) symmetry. When considering one-dimensional (1D) non-Hermitian systems, from the symmetry point of view, $\mathcal{P}$ symmetry and NHSE cannot coexist. It indicts that $\mathcal{P}$ symmetry can be utilized as a criterion to determine whether the NHSE can occur or not. However,  in the most 1D non-Hermitian
systems \cite{Lee2016,Yao2018,Yao2018a,Song2019,Kunst2018,Li2020,Feng2024,Li2024}, the $\mathcal{P}$ symmetry is prohibited. To solve this problem, here we consider enlarging the symmetry group and utilizing the combined-$\mathcal{P}$ symmetry \cite{Luders1954,Lande1957,Landau1957}
instead. We first prove a theorem that the combined-$\mathcal{P}$ symmetry can be used as a criterion in 1D to determine whether the NHSE can be present or not. Then we take the 1D non-Hermitian Kitaev model as an example, to show that the combined-$\mathcal{P}$ symmetry, i.e., $\mathcal{PC}$ symmetry with $\mathcal{C}$ referring to the particle-hole symmetry (charge conjugation), can determine the fate of NHSE. It is verified by both a mathematical proof constructed through the non-Bloch theory and the exact diagonalization numerical studies.

\textit{Theorem $\raisebox{0.01mm}{---}$} For a 1D non-Hermitian lattice system captured by the Hamiltonian $\mathbf{H}$, where the Hamiltonian matrix can not be block-diagonalized under any permutations, through the non-Bloch band theory the bulk mode can be expressed as $|\psi \rangle =\sum\limits_{j,i}(\beta_{j})^{i}\left\vert \phi _{j}\right\rangle \otimes\left\vert i\right\rangle $, with $\beta _{j}$ being the non-Bloch complex wave vector \cite{Yao2018} and $\left\vert \phi _{j}\right\rangle $ representing the corresponding non-spatial eigenstate. $i$ labels the lattice site. A combined-$\mathcal{P}$ symmetry, i.e., $\mathcal{S} \equiv \mathcal{PX}$, where $\mathcal{X}$ stands for a non-spatial symmetry, can be defined. We then require $\mathcal{S}$ to satisfy the following conditions: \\
 ({\romannumeral1}) $\mathcal{S}$ commutes with $\mathbf{H}$, i.e.,
\begin{equation}
[\mathbf{H}, \mathcal{S}] = 0.  \label{Eq.commutationrelay}
\end{equation}
({\romannumeral2})The bulk mode under the transformation of $\mathcal{S}$ satisfies the following relation
\begin{equation}
\mathcal{S}|\psi \rangle =\sum\limits_{j,i}(\beta_{j})^{i}\left\vert \phi _{j}^{\prime }\right\rangle \otimes \left\vert-i\right\rangle,  \label{Eq.SPhitransf}
\end{equation}
then we can conclude:

\textit{If $\mathcal{S}$ symmetry exists in the considered 1D non-Hermitian lattice system, NHSE will disappear, otherwise NHSE will be present.}

\textit{Proof of the theorem $\raisebox{0.01mm}{---}$} The proof of the above theorem is quite straightforward. First, we reexpress condition ({\romannumeral1}) as
\begin{equation}
\mathcal{S}\mathbf{H}\mathcal{S}^{-1}=\mathbf{H},  \label{Eq.combsymmetry}
\end{equation}
Then, suppose there is a bulk mode under OBC, which satisfies the relation $\mathbf{H}\left\vert \Psi _{1}\right\rangle=E\left\vert \Psi _{1}\right\rangle$, where $E$ and $\left\vert \Psi _{1}\right\rangle$ are the corresponding eigenenergy and eigenstate, respectively. Using Eq.~(\ref{Eq.combsymmetry}), we can obtain
\begin{equation}
\mathbf{H}\mathcal{S}^{-1}\left\vert \Psi _{1}\right\rangle=E\mathcal{S}^{-1}\left\vert \Psi _{1}\right\rangle.
\label{Phi2E}
\end{equation}%
Therefore, $\left\vert \Psi _{2}\right\rangle\equiv\mathcal{S}^{-1}\left\vert \Psi _{1}\right\rangle$ is also a bulk eigenstate with the same eigenenergy $E$ as $\left\vert \Psi _{1}\right\rangle$. From the non-Bloch band theory, it is shown that bulk states with the same eigenenergy have identical non-Bloch complex wave vectors $\beta$ from the same generalized Brillouin zone (GBZ) \cite{Song2019,Yokomizo2019}. When $\left\vert \beta\right\vert=1$, bulk states are delocalized and no skin effect occurs. While $\left\vert \beta\right\vert\neq 1$, skin effect appears and bulk states with the same eigenenergy will be localized at the same end of the 1D chain.

Starting from Eq. (\ref{Phi2E}), we can obtain that $\left\vert \Psi _{1}\right\rangle=\mathcal{S}\left\vert \Psi _{2}\right\rangle$. If the skin effect exists and without loss of generality, assuming the bulk state $\left\vert \Psi _{1}\right\rangle$ to be localized at one end of the 1D chain, through using condition ({\romannumeral2}), it is shown that $\left\vert \Psi _{2}\right\rangle$ should be localized at the opposite end of the chain. However, this conclusion is obviously contradictory to the results obtained from the non-Bloch band theory that bulk states with the same eigenenergy will be localized at the same end of the 1D chain. Therefore, if $\mathcal{S}$ exists in the considered 1D non-Hermitian lattice system, NHSE will disappear, otherwise NHSE will be present and the theorem is proved.

\textit{1D non-Hermitian Kitaev chain $\raisebox{0.01mm}{---}$} As an example, we consider a 1D non-Hermitian Kitaev chain with asymmetric hopping, which can be captured by the following Hamiltonian
\begin{equation}
\mathbf{H}=\sum\limits_{n=1}^{L}[-(t+\frac{\gamma }{2})\hat{c}_{n}^{\dagger }%
\hat{c}_{n+1}-(t-\frac{\gamma }{2})\hat{c}_{n+1}^{\dagger }\hat{c}%
_{n}+\Delta (\hat{c}_{n}\hat{c}_{n+1}+h.c.)],  \label{Eq.Hamiltonian}
\end{equation}
where $\hat{c}_{n}^{\dagger}(\hat {c}_{n})$ is the fermionic creation (annihilation) operator. $t$
and $\gamma$ describe the symmetric and asymmetric hopping amplitude between nearest neighbor sites, respectively. $L$ is the total lattice site. $\Delta$ represents the $p$-wave pairing in the 1D non-Hermitian Kitaev chain, which is taken as real numbers.
It is worthy to note that the model Hamiltonian in Eq.~(\ref{Eq.Hamiltonian}) preserves the combined-$\mathcal{P}$ symmetry, i.e., $\mathcal{S}\equiv \mathcal{PC}$, where $\mathcal{C}$ represents the particle-hole (charge conjugation) symmetry and $\mathcal{P}$ stands for the spatial reflection \cite{Cai2021,Han2020,Buca2012,Xiao2018,Zirnbauer2021}. We can further express $\mathcal{S}$ in the following matrix form
\begin{equation}
S =\sigma _{y}\otimes R,
\end{equation}
with $\sigma _{y}$ being the Pauli matrix. $R$ is a $L\times L$ matrix with $R_{nn^{\prime}}=(-1)^{n}\delta _{n^{\prime},L-n+1}$. It is straightforward to prove that $\mathcal{S}\mathbf{H}\mathcal{S}^{-1}=\mathbf{H}$. Therefore, through our proved theorem above, we can conclude that NHSE will disappear in the 1D
non-Hermitian system captured by Eq.~(\ref{Eq.Hamiltonian}).
\begin{figure*}[!ht]
\centering
\includegraphics[width=0.8\textwidth]{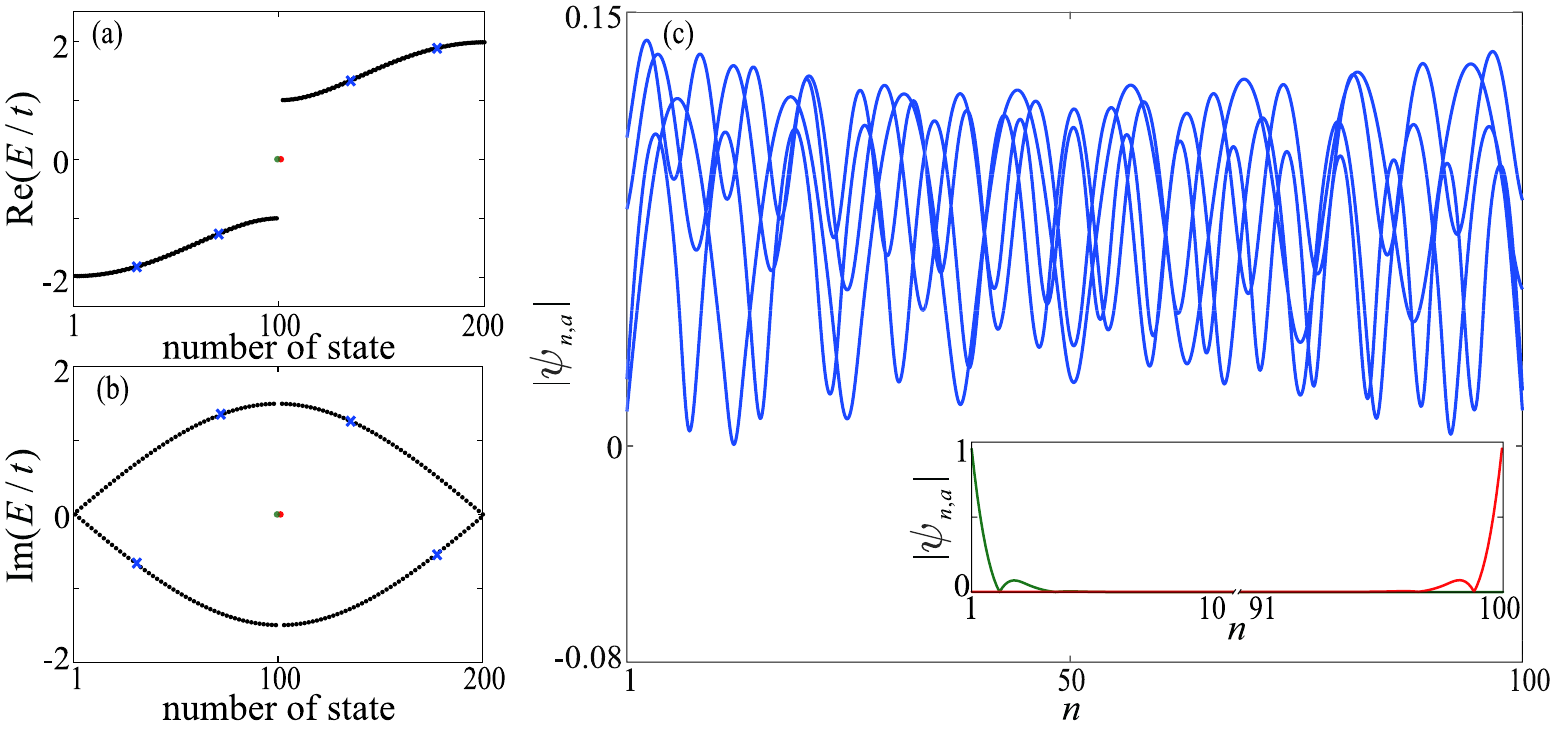}
\caption{(a) and (b) Energy spectrum of the model of Hamiltonian in Eq. (\ref{Eq.Hamiltonian}) under OBC.  The real and imaginary part of the complex eigenenergy spectrum are shown in (a) and (b), respectively. (c) Profiles of the chosen eigenstates. The eigenstates of the bulk modes marked by blue crosses in (a) and (b) are delocalized at the 1D system, indicating the disappearance of NHSE. The inset shows the eigenstates of the topological edge modes, marked by green and red dots in (a) and (b). Other parameters are chosen as $L=100$, $\Delta/t=0.5$, $\gamma/t=3/2$.}
\label{figure:1}
\end{figure*}

To verify it, we numerically solve the eigen-problem associated with $\mathbf{H}$ in
Eq.~(\ref{Eq.Hamiltonian}) under OBC through the following diagonalization procedure. Through introducing the Nambu spinors $(\hat{c}_1, \hat{c}_2, ..., \hat{c}_L, \hat{c}_1^\dagger, \hat{c}_2^\dagger, ..., \hat{c}_L^\dagger)$, the system in
Eq.~(\ref{Eq.Hamiltonian}) can be captured by the following Bogoliubov-de Gennes (BdG) Hamiltonian
\begin{equation}
\mathcal{H}_{BdG}=\left(
\begin{array}{cc}
\mathbf{h} & \mathbf{\Delta}\\
-\mathbf{\Delta} & -\mathbf{h}^{\dagger }%
\end{array}%
\right),  \label{Eq.bdgmatrix}
\end{equation}
with $\mathbf{h}_{nn^{\prime}}=-(t+\frac{\gamma}{2})\delta_{n^{\prime},n+1}-(t-\frac{\gamma}{2})\delta_{n^{\prime},n-1}$, $\mathbf{\Delta}_{nn^{\prime}}=-\Delta(\delta_{n^{\prime},n+1}-\delta
_{n^{\prime},n-1}) $. The eigenstates and corresponding eigenenergies can be obtained by diagonalizing Eq.~(\ref{Eq.bdgmatrix}). As shown in Fig.~\ref{figure:1}, the energy spectrum under OBC is obtained. The real
and imaginary part of that are shown in Fig. \ref{figure:1}(a) and (b), respectively. Without loss of generality, we randomly select four bulk states marked by blue cross and display their corresponding wave functions in Fig. \ref{figure:1}(c). It is shown that in the 1D non-Hermitian system captured by Eq. (\ref{Eq.Hamiltonian}), the NHSE is anomalously blocked by the presence of the
combined-$\mathcal{P}$ symmetry, i.e., $\mathcal{PC}$ symmetry here.

\textit{Analysis from the non-Bloch theory $\raisebox{0.01mm}{---}$} Besides the numerical simulations, in the following we employ the non-Bloch band theory \cite{Yao2018,Yao2018a,Yokomizo2019,Song2019} to study the NHSE in the system described by Eq. (\ref{Eq.Hamiltonian}). Starting from Eq. (\ref{Eq.bdgmatrix}), assuming the eigenstate of $\mathcal{H}_{BdG}$ can be expressed as $|\psi \rangle=(\psi _{1,a},\psi _{2,a},...,\psi _{L,a},\psi
_{1,b},\psi _{2,b},...,\psi _{L,b})^T$, we can derive the real-space eigenequation
\begin{eqnarray}
E\psi _{n,a} &=&-(t-\frac{\gamma }{2})\psi _{n-1,a}-(t+\frac{\gamma }{2}%
)\psi _{n+1,a}  \notag \\
&&+\Delta (\psi _{n-1,b}-\psi _{n+1,b}),  \notag \\
E\psi _{n,b} &=&(t+\frac{\gamma }{2})\psi _{n-1,b}+(t-\frac{\gamma }{2})\psi
_{n+1,b}  \notag \\
&&+\Delta (\psi _{n+1,a}-\psi _{n-1,a}).  \label{Eq.eigenequation}
\end{eqnarray}%
We then further take an ansatz for the wave function as a linear combination $\psi _{n,\mu
}=\sum_{j}\phi _{n,\mu }^{(j)}$, with $\mu =a,b$. Each $\phi _{n,\mu }^{(j)}$ takes the exponential form $(\phi
_{n,a}^{(j)},\phi _{n,b}^{(j)})=\beta _{j}^{n}(\phi _{a}^{(j)},\phi
_{b}^{(j)})$, where $\phi _{\mu }^{(j)}$ is the eigenstate with eigenvalue $\beta
_{j}$ determined by the following eigenvalue equation
\begin{equation}
\det [\mathcal{H}(\beta )-E]=0,
\end{equation}%
with $\mathcal{H}(\beta )$ being defined as
\begin{equation}
\mathcal{H}(\beta )=\frac{\gamma }{2}(\beta ^{-1}-\beta )\mathbb{I}-t(\beta
^{-1}+\beta )\sigma _{z}+i\Delta (\beta ^{-1}-\beta )\sigma _{y},
\label{Eq.Hbetamatr}
\end{equation}%
where $\mathbb{I}$ is the identity matrix, $\sigma _{y,z}$ are the Pauli matrices. The spectrum of the system can be determined through the following relation
\begin{equation}
(\Delta ^{2}-t^{2}+\frac{\gamma ^{2}}{4})(\beta ^{2}+\beta ^{-2}-2)+E\gamma
(\beta -\beta ^{-1})+E^{2}-4t^{2}=0.
\label{Eq.ChPbeta}
\end{equation}
To rewrite Eq. (\ref{Eq.ChPbeta}), we define $x \equiv\beta -\beta ^{-1}$, Eq. (\ref{Eq.ChPbeta}) can thus be expressed as
\begin{equation}
(\Delta ^{2}-t^{2}+\frac{\gamma ^{2}}{4})x^{2}+E\gamma x+(E^{2}-4t^{2})=0.
\label{Eq.quarticeq}
\end{equation}%
There are two solutions of $x$ obtained from Eq. (\ref{Eq.quarticeq}) $x_{\pm}=\frac{-E\gamma \pm \sqrt{%
E^{2}\gamma ^{2}-4(\Delta ^{2}-t^{2}+\frac{\gamma ^{2}}{4})(E^{2}-4t^{2})}}{%
2(\Delta ^{2}-t^{2}+\frac{\gamma ^{2}}{4})}$. Then $\beta$ can be solved through the following relation
\begin{equation}
\beta ^{2}-x_{\pm}\beta -1=0.
\label{quadraticeq}
\end{equation}
The solution of $\beta$ contains four elements $\beta_j=\beta _{m_\pm}$ $(m=1,2)$. From Eq.~(\ref{quadraticeq}), $\beta$ should satisfy the following relation
\begin{align}
\beta _{1_+}\times \beta _{2_+}& =-1,  \notag \\
\beta _{1_-}\times \beta _{2_-}& =-1. \label{Vietatheorem}
\end{align}%
Without loss of generality, let us assume that $\beta$ satisfies the conditions that $\left\vert \beta _{1_+}\right\vert \leq 1 \leq \left\vert \beta
_{2_+}\right\vert $ and $\left\vert \beta _{1_-}\right\vert \leq 1 \leq \left\vert \beta_{2_-}\right\vert $. From the non-Bloch band theory \cite{Yao2018,Yao2018a,Yokomizo2019,Song2019}, the condition to get the continuum bands can be written as
\begin{equation}
\left\vert \beta _{1_+}\right\vert \leq \left\vert \beta
_{1_-}\right\vert =\left\vert \beta _{2_-}\right\vert \leq \left\vert
\beta _{2_+}\right\vert,\label{betaup1}
\end{equation}
or
\begin{equation}
\left\vert \beta _{1_-}\right\vert \leq \left\vert \beta
_{1_+}\right\vert =\left\vert \beta _{2_+}\right\vert \leq \left\vert
\beta _{2_-}\right\vert.\label{betaup2}
\end{equation}
Combining Eq.~(\ref{betaup1})/(\ref{betaup2}) with Eq.(\ref{Vietatheorem}), we can get that $\beta$ constituting the GBZ should satisfy the relation $\left\vert \beta \right\vert=1$ (See details in the Supplemental Material (SM)). Since $\left\vert \beta\right\vert=1$, the bulk modes will persist in the extended states \cite{Yao2018,Yokomizo2019, Li2020} and NHSE will disappear.

\begin{figure}[!htbp]
\centering
\includegraphics[width=0.5\textwidth]{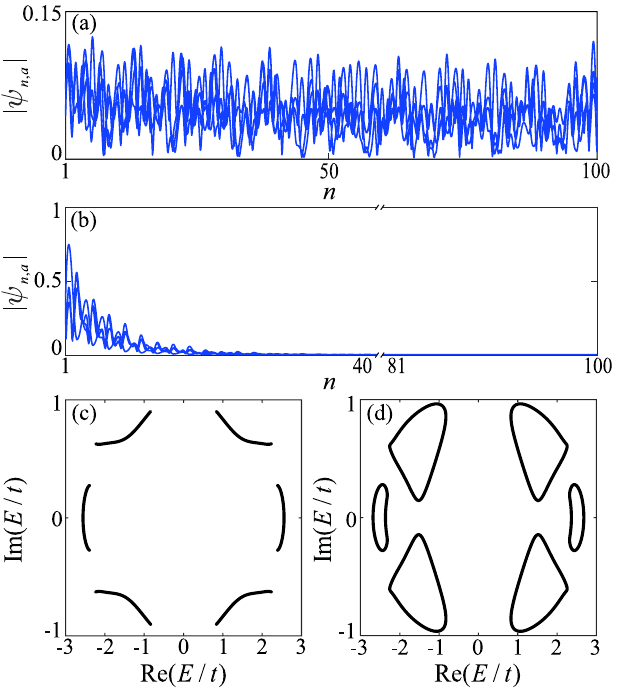}
\caption{ (a) and (b) Profiles of the bulk eigenstates when considering adding $\mathbf{H}^\prime$ to $\mathbf{H}$ in Eq.~(\ref{Eq.Hamiltonian}), which can turn on (off) the $\mathcal{PC}$ symmetry of the system. In (a), $\theta=0$, $\mathcal{PC}$ symmetry
is persevered, NHSE disappears and the bulk modes are delocalized. In (b), $\theta=\pi/4$, $\mathcal{PC}$ symmetry is broken, NHSE is restored and the accumulation of eigen-wavefunctions occurs. (c) and (d) Complex spectra under the periodic boundary condition in the complex-energy plane corresponding to the case in (a) and (b), respectively. Here $V/t=2$ and other parameters are same as Fig. \ref{figure:1}.}
\label{figure:2}
\end{figure}

\textit{Discussion} \& \textit{Conclusion} $\raisebox{0.01mm}{---}$ To show the generality of our developed symmetry-based method to examine the origin/reason for the occurrence of NHSE, we extend the considered model in Eq. (\ref{Eq.Hamiltonian}) by adding an additional term $\mathbf{H}^\prime$, where the combined-$\mathcal{P}$ symmetry can be turned on or off through tuning the Hamiltonian parameters. For instance, let us consider setting $\mathbf{H}^\prime=\sum\limits_{n}V_n\hat{c}_{n}^{\dagger }\hat{c}_{n}$, where $V_n=V\sin (2\pi n/3 +\theta)$ with $V$ describing the strength of the potential and $\theta\in [0, 2\pi)$. When $\theta$ being an integral multiple of $\pi/3$, the system will maintain the combined-$\mathcal{P}$ symmetry, i.e., $\mathcal{PC}$ symmetry defined above. While $\theta$ is chosen as other angles, the $\mathcal{PC}$ symmetry of the system will be broken. As shown in Fig. \ref{figure:2}(a), when $\mathcal{PC}$ symmetry
is persevered  in the system ($\theta=0$), NHSE will not occur and the bulk modes of the system will remain be extended, which is consistent with our proposed theorem. While $\mathcal{PC}$ symmetry is broken, as shown in Fig. \ref{figure:2}(b) ($\theta=\pi/4$), NHSE is restored agreeing with our proposed theorem.

Besides the bulk modes, as shown in Fig. \ref{figure:1}, there are two topological edge modes (marked with colored dots Fig. \ref{figure:1}(a) and (b)). The corresponding wavefunctions are shown in the inset of Fig. \ref{figure:1}(c), where the two edge states are localized at the opposite ends of the 1D chain. The existence of these topological edge states \cite{Hatsugai1993,Hasan2010,Qi2011} can be understood through the bulk-edge correspondence principle via analysing the non-Bloch Zak phase \cite{Bartlett2021,Zak1989}, which is defined as $\gamma _{\pm }=i\oint_{C_{\beta }}d\beta \left\langle u_{\pm,L}\right\vert
\partial _{\beta }\left\vert u_{\pm,R}\right\rangle$, where $\left\vert u_{\pm ,R}\right\rangle $ and $\left\langle u_{\pm ,L}\right\vert $ are the right and left eigenvectors with the eigenenergies ${E_{\pm }}$ of Eq. (\ref{Eq.Hbetamatr}). It is shown that the accumulated Zak phase when moving across the GBZ loop is equal to $\pm\pi$, indicating that the system is topologically non-trivial and there are topological edge modes at the end of chain when considering OBC.

A symmetry-based method to examine the origin/reason for the occurrence of NHSE has been developed in this work. We prove a theorem that the combined-$\mathcal{P}$ symmetry can be used as a criterion in 1D to determine whether the NHSE can be present or not. Taking the 1D non-Hermitian Kitaev model as an example, through both a mathematical proof via the non-Bloch band theory and the exact diagonalization numerical studies, we verify our proposed theorem.  This result reveals a profound connection between the symmetry and the fate of NHSE. This approach is rather generic to other 1D non-Hermitian systems \cite{Zhao2021, Yokomizo2021} than restricted to the case considered in this work.
Our results may also find their usage in a variety of future experiments, including in  topolectric circuit, photonic lattice, quantum gases, as well as the solid-state materials.

\textit{Acknowledgment $\raisebox{0.01mm}{---}$}
This work is supported by the National Key R$\&$D Program of China (2021YFA1401700), NSFC (Grants No. 12074305, 12147137, 11774282), the Fundamental Research Funds for the Central Universities (Grant No. xtr052023002), the Shaanxi Fundamental Science Research Project for Mathematics and Physics (Grant No. 23JSZ003) and  Xiaomi Young Scholar Program. We also thank the HPC platform of Xi'an Jiaotong University, where our numerical calculations was performed.

\bibliographystyle{apsrev}
\bibliography{blockreferrev2}

\onecolumngrid

\renewcommand{\thesection}{S-\arabic{section}}
\setcounter{section}{0}  
\renewcommand{\theequation}{S\arabic{equation}}
\setcounter{equation}{0}  
\renewcommand{\thefigure}{S\arabic{figure}}
\setcounter{figure}{0}  

\indent

\begin{center}\large
\textbf{Supplementary Material}
\end{center}

\section{Derivation of the condition for the continuum bands}
In this section, we derive the condition for continuum bands of the model Hamiltonian in Eq. (5). To this end, we focus on boundary conditions in a finite open chain with $L$ sites. When further considering the real space eigenequation Eq. (8) at the left end of the open chain around $n = 1$ and at the right end of the open chain around $n = L$, we obtain the following relations
\begin{eqnarray}
E\psi _{1,a} &=&-(t+\frac{\gamma }{2})\psi _{2,a}-\Delta \psi _{2,b},  \notag
\\
E\psi _{1,b} &=&(t-\frac{\gamma }{2})\psi _{2,b}+\Delta \psi _{2,a},  \notag
\\
E\psi _{L,a} &=&-(t-\frac{\gamma }{2})\psi _{L-1,a}+\Delta \psi _{L-1,b},
\notag \\
E\psi _{L,b} &=&(t+\frac{\gamma }{2})\psi _{L-1,b}-\Delta \psi _{L-1,a}.\label{Eq.S1}
\end{eqnarray}
From the real space eigenequation Eq. (8), one can easily find that the ratio between the values of $\phi _{n,\mu }^{(j)}$ is fixed through the relation
\begin{eqnarray}
\phi _{a}^{(j)} &=&\frac{\Delta (\beta _{j}^{-1}-\beta _{j})}{E+(t-\frac{%
\gamma }{2})\beta _{j}^{-1}+(t+\frac{\gamma }{2})\beta _{j}}\phi _{b}^{(j)},
\notag \\
\phi _{b}^{(j)} &=&\frac{\Delta (\beta _{j}-\beta _{j}^{-1})}{E-(t+\frac{%
\gamma }{2})\beta _{j}^{-1}-(t-\frac{\gamma }{2})\beta _{j}}\phi _{a}^{(j)}.\label{Eq.S2}
\end{eqnarray}
where $\beta_j=\beta _{m_\pm}$ $(m=1,2)$ (as defined in the main text). Through substituting Eq. (\ref{Eq.S2}) into Eq. (S1), one can reduce the problem into four linear equations for $\phi_{a}^{(j)}$ and obtain
\begin{eqnarray}
\sum_{j}A_{j}\phi _{a}^{(j)} &=&0,  \notag \\
\sum_{j}B_{j}\phi _{a}^{(j)} &=&0,  \notag \\
\sum_{j}C_{j}\beta _{j}^{L-1}\phi _{a}^{(j)} &=&0,  \notag \\
\sum_{j}D_{j}\beta _{j}^{L}\phi _{a}^{(j)} &=&0.  \label{Eq.S3}
\end{eqnarray}%
where $A_{j}$,$B_{j}$,$C_{j}$and $D_{j}$are defined as%
\begin{eqnarray}
A_{j} &=&E\beta _{j}+(t+\frac{\gamma }{2})\beta _{j}^{2}+\frac{\Delta
^{2}(\beta _{j}^{2}-1)}{E\beta _{j}^{-1}-(t+\frac{\gamma }{2})\beta
_{j}^{-2}-(t-\frac{\gamma }{2})},  \notag \\
B_{j} &=&\frac{\Delta (\beta _{j}-\beta _{j}^{-1})[E-(t-\frac{\gamma }{2}%
)\beta _{j}^{2}]}{E-(t+\frac{\gamma }{2})\beta _{j}^{-1}-(t-\frac{\gamma }{2}%
)\beta _{j}}-\Delta \beta _{j}^{2},  \notag \\
C_{j} &=&t-\frac{\gamma }{2}+E\beta _{j}-\frac{\Delta (\beta _{j}-\beta
_{j}^{-1})}{E-(t+\frac{\gamma }{2})\beta _{j}^{-1}-(t-\frac{\gamma }{2}%
)\beta _{j}},  \notag \\
D_{j} &=&\frac{1-\Delta (t+\frac{\gamma }{2})(1-\beta
_{j}^{-2})}{E-(t+\frac{\gamma }{2})\beta _{j}^{-1}-(t-\frac{\gamma }{2}%
)\beta _{j}}.
\end{eqnarray}%
From Eq. (\ref{Eq.S3}), we can obtain the condition for the existence of
nontrivial solutions for $\phi _{a}^{(j)}$as
\begin{equation}
\begin{vmatrix}
A_{1_{+}} & A_{1_{-}} & A_{2_{-}} & A_{2_{+}} \\
B_{1_{+}} & B_{1_{-}} & B_{2_{-}} & B_{2_{+}} \\
C_{1_{+}}(\beta _{1_{+}})^{L-1} & C_{1_{-}}(\beta _{1_{-}})^{L-1} &
C_{2_{-}}(\beta _{2_{-}})^{L-1} & C_{2_{+}}(\beta _{2_{+}})^{L-1} \\
D_{1_{+}}(\beta _{1_{+}})^{L} & D_{1_{-}}(\beta _{1_{-}})^{L} &
D_{2_{-}}(\beta _{2_{-}})^{L} & D_{2_{+}}(\beta _{2_{+}})^{L}%
\end{vmatrix}%
=0.  \label{Eq.S4}
\end{equation}%
The determinant in Eq. (\ref{Eq.S4}) can be expressed as an algebraic
equation for $\beta _{j}$
\begin{equation}
\frac{1}{2}\sum\limits_{\sigma }\mathrm{sgn}(\sigma )(A_{\sigma
(1)}B_{\sigma (2)}-A_{\sigma (2)}B_{\sigma (1)})C_{\sigma (3)}D_{\sigma
(4)}\beta _{\sigma (3)}^{L-1}\beta _{\sigma (4)}^{L}=0,  \label{Eq.S5}
\end{equation}%
where the sum is taken over all the permutations $\sigma $ for four objects.

Starting from Eq. (14), without loss of generality, let us assume that $ \beta_{j}$ satisfies the following conditions $\left\vert \beta _{1_+}\right\vert \leq 1 \leq \left\vert \beta_{2_+}\right\vert $ and $\left\vert \beta _{1_-}\right\vert \leq 1 \leq \left\vert \beta_{2_- }\right\vert $. Therefore, we can number these four $ \beta_{j}$ so as to satisfy the following two possible orders: ({\romannumeral1}) $\left\vert \beta _{1_+}\right\vert \leq \left\vert \beta_{1_-}\right\vert \leq\left\vert \beta _{2_-}\right\vert \leq \left\vert\beta_{2_+}\right\vert$, ({\romannumeral2}) $\left\vert \beta
_{1_-}\right\vert \leq \left\vert \beta _{1_+}\right\vert \leq
\left\vert \beta _{2_+}\right\vert \leq \left\vert \beta_{2_-}\right\vert $. Let us first consider case ({\romannumeral1}), when the solutions of Eq. (\ref{Eq.S5}) are densely distributed for a large $L$, if $\left\vert \beta _{1_-}\right\vert \neq \left\vert \beta_{2_-}\right\vert $, in the thermodynamic limit $L\rightarrow \infty$, the term which is proportional to $(\beta _{2_-}\beta _{2_+})^{L-1}$ becomes the leading term in Eq. (\ref{Eq.S5}). Therefore, to satisfy Eq. (\ref{Eq.S5}), we obtain the following relation $(A_{1_{+}}B_{1_{-}}-A_{1_{-}}B_{1_{+}})(C_{2_{-}}D_{2_{+}}\beta
_{2_{+}}-C_{2_{+}}D_{2_{-}}\beta _{2_{-}})=0 $ in the thermodynamic limit. Combining the above relation with the Eq. (11), it is shown that the eigenenergies obtained are restricted to discrete values and cannot represent continuum bands. Conversely, when $\left\vert \beta _{1_-}\right\vert =\left\vert \beta _{2_-}\right\vert $, there are two leading terms in Eq. (\ref{Eq.S5}), which are proportional to $(\beta _{2_-}\beta _{2_+})^{L-1}$ and $(\beta _{1_{-}}\beta _{2_+})^{L-1}$, respectively, when considering $L\rightarrow \infty$. Then, from Eq. (\ref{Eq.S5}) we can obtain $(\frac{\beta_{1_{-}}}{\beta _{2_{-}}})^{L}=\frac{(A_{1_{+}}B_{1_{-}}-A_{1_{-}}B_{1_{+}})(C_{2_{-}}D_{2_{+}}\beta _{2_{+}}-C_{2_{+}}D_{2_{-}}\beta_{2_{-}})}{(A_{1_{+}}B_{2_{-}}-A_{2_{-}}B_{1_{+}})(C_{1_{-}}D_{2_{+}}\beta_{2_{+}}-C_{2_{+}}D_{1_{-}}\beta _{1_{-}})}$. The above relation allows a dense set of solutions when the relative phase between $\left\vert \beta
_{1_{-}}\right\vert $ and $\left\vert \beta _{2_{-}}\right\vert $ is continuously varied. Therefore, $\left\vert \beta _{1_{-}}\right\vert =\left\vert \beta _{2_{-}}\right\vert $ emerges as an appropriate condition for continuum bands. For case ({\romannumeral2}), a similar analysis can be applied and  the condition for continuum bands is obtained as $\left\vert \beta _{1_+}\right\vert =\left\vert \beta _{2_+}\right\vert $. Therefore, Eq. (15) and Eq. (16) in the main text can be got.

\end{document}